\documentstyle[aps]{revtex}
\begin{document}
\twocolumn
\draft              
\title{Generalized $t$-$t^\prime$-$J$ model: parameters and 
single-particle spectrum for electrons and holes in copper oxides.}
\author{V. I. Belinicher, A. L. Chernyshev, and V. A. Shubin}
\address{Institute of Semiconductor Physics, 630090, Novosibirsk, Russia}
\date{\today}
\maketitle
\begin{abstract}
 A microscopically based Hamiltonian of the generalized 
$t$-$t^\prime$-$J$ model is presented. Two types of the additional 
$t^\prime$-terms are discussed. The numerical range of the 
amplitudes corresponding to the
additional $t^\prime$-terms for the real CuO$_2$ planes is derived 
from the three-band model calculations. Using the variational 
spin-polaron approach the single-carrier dispersions in the 
generalized $t$-$t^\prime$-$J$ model are calculated both for the 
hole- and electron-doped systems. The hole and  electron band minima 
are found to be at points $(\pm\pi/2,\pm\pi/2)$,$(0,\pm\pi)$,  and  
$(\pm\pi,0)$,  respectively. The band minima shifts 
$|\Delta_{(0,\pi)-(\pi/2,\pi/2)}|$ are not small ($\sim J$). 
The bandwidths for both cases of doping  are  found to be $1.5-4.0$ 
times larger than those in the $t$-$J$ model.  
\vskip 0.5cm 
\noindent 
\end{abstract}
\narrowtext

\section{Introduction}
\label{sec:level1}


There is a general agreement that the three-band Hubbard model is an  
appropriate basis for consideration of the spin and charge 
excitations in the CuO$_2$ planes of high-temperature 
superconductors \cite{Em1,Abrah1,Lok1}.  One of the interesting 
problems for these systems is the hole energy spectrum. One-hole 
energy calculation in the framework of the above mentioned model has 
been done by Barabanov, Kuzian, Maksimov, and Uimin \cite{Ba1,Ba2} 
who used the variational approach. This method seems to be too  
complicated since the characteristic energy scale of the three-band 
model is a few electron-volts while that of 
the hole energy spectrum is several tenths of an electron-volt.  
Therefore, it is natural to obtain first the low-energy limit of 
the three-band model \cite{Zh1,Esk1,Jeff1,Alig1,Yush1,Bel1,Bel2} and 
then to investigate the spin and charge degrees of freedom in the 
framework of this effective model \cite{Dag1,Sush1,Chub1,Good1,Toh1}.

It is widely accepted by now that the simple model which contains in 
itself the interacting spin and charge degrees of freedom is the so 
called $t$-$J$ model
\begin{eqnarray}
\label{1}
H_{t-J}= t \sum_{\langle ij\rangle,\alpha}
\tilde{c}^{\dag}_{i,\alpha}\tilde{c}_{j,\alpha} + J \sum_{\langle ij\rangle}
{\bf S}_i{\bf S}_j  ,
\end{eqnarray}
where $\langle ij \rangle$ denotes the nearest neighbor sites, ${\bf 
S}_i$ is the local spin operator, 
$\tilde{c}^{\dag}_{i,\alpha}(\tilde{c}_{i,\alpha})$ is the 
constrained fermion creation (annihilation) operator. The explicit 
form of this constraint depends on the type of doping.

The considerable attention that this model has received
originates partly from the fact that it can be derived from the above 
mentioned three-band or more general $d-p$ models which take 
into account the detailed electronic structure of the copper oxides. 
Recently it has been shown \cite{Bel4} that this derivation can be 
done quantitatively for the real CuO$_2$ planes by the use of an 
additional calculation of some experimentally observable values.

The behavior of a single quasiparticle in the $t$-$J$ model has 
been studied intensively using both analytical and numerical 
techniques \cite{Sush1,Chub1,Auer1,Dagrev}. These investigations have 
clearly shown that the naive tight-binding picture is completely 
inadequate for the carrier motion on the antiferromagnetic 
background. Namely, for the realistic $t/J\simeq 2-3$  the bandwidth 
is of the order of $2J$, not $W_0=8t$. This strong (of the order of 
10 times) bandwidth suppression results from the distortion of the 
spin background by the carrier hopping from one sublattice to the 
other. In this situation, when the characteristic energy scale 
becomes $J$, even small (compared to $t$) single-sublattice hopping 
$t^{\prime}$ can be the key parameter for the subtle details of the 
energy spectrum and other features of the doped systems. This is 
clear since the single-sublattice motion does not disturb the 
spin background. Hence, a careful analysis of the low-energy single 
band limit of the realistic $d-p$ models should be done keeping all 
essential terms over the $t^{\prime}/J$ (not $t^{\prime}/t$) 
parameter.

The simplest form of the additional $t^{\prime}$-terms can be written 
as
\begin{eqnarray}
\label{1a}
H_{t^{\prime}}= t^{\prime} \sum_{\langle ij\rangle_2,\alpha}
\tilde{c}^{\dag}_{i,\alpha}\tilde{c}_{j,\alpha} ,
\end{eqnarray}
emphasizing that the nonzero O-O hopping provides large enough 
transfer amplitude to the next-nearest ($\langle ij \rangle_2$) 
neighbor \cite{Dagrev,Lee1}. This term (\ref{1a}) alone has been 
found to  be responsible for the CuO$_2$ plane electron-hole 
asymmetry and to be useful for the interpretation of the recent 
angle-resolved photoemission experiments \cite{Naz1}.

The goal of the present paper is to exhibit the {\it general } form 
of the additional $t^{\prime}$-terms which one can get from the 
three-band Hubbard or the other first-principle models, to determine 
numerical ranges of all essential parameters for the real CuO$_2$ 
planes, and to calculate a single-particle dispersion using the 
spin-polaron ideas.  Namely, we will demonstrate that two types 
of the additional $t^{\prime}$-terms naturally originate in the 
single-band model:  (i) the terms arising from the O-O hopping, and 
(ii) the second-order high-energy channels terms 
\cite{Hy1,Psal1,Oles1}. The importance of both of them will be 
clearly shown. The first ones are the key parameters for the band 
minima shifts, while the second ones lead to the growing of the 
bandwidth. A quite similar model was investigated in the mean-field 
approximation for a special range of parameters in the work by 
Onufrieva {\it et al} \cite{Onuf1}. This approximation neglects some 
essential features of the hole (electron)-spin interaction and thus 
has only a qualitative character. 

Single-hole energy calculation in the limiting case of the three-band 
model using the self-consistent Born approximation has been done 
in the work by Starykh {\it et al} \cite{Star1}.  In spite of 
 neglecting some features of the problem, the results of this 
approach are in a qualitative agreement with the present paper.

The quantitative reduction of the three-band model developed in 
Ref.\cite{Bel4} for the real CuO$_2$ system provides the method of 
the derivation of a set of the {\it realistic} ranges for all 
additional $t^{\prime}$-terms. We will show that the band minima 
shifts as well as the bandwidths are not small for these ranges of 
parameters.  Also, the density of states (DOS) characteristic 
features will be discussed.

The paper is organized as follows. In Sec. II we give the general 
form of the corrections to the Hamiltonian (\ref{1}), discuss them, 
and show the possible parameters ranges. In Sec. III we discuss our 
results for the spin-polaron dispersions and for the DOS features. 
Finally, in Sec. IV, we draw conclusions.

\section {Realistic low-energy model}


The most accurate form of the effective $t^{\prime}$-terms that 
follows from the three-band Hubbard model is \cite{Bel4,Bel3}:
\begin{eqnarray}
\label{2}
&&H_{t^{\prime}}=t^{\prime}\sum_{\langle ij 
\rangle_{2},\alpha}\tilde{c}^{\dag}_{i,\alpha}\tilde{c}_{j,\alpha}
+t^{\prime \prime}\sum_{\langle ij 
\rangle_{3},\alpha}\tilde{c}^{\dag}_{i,\alpha}\tilde{c}_{j,\alpha}
\nonumber\\
&&\phantom{H_{t^{\prime}}=} \mbox{}+t^{N}\sum_{\langle ilj 
\rangle,\alpha}\tilde{c}^{\dag}_{i,\alpha}\tilde{c}_{j,\alpha}\hat{N}_l
+t^{S}\sum_{\langle ilj 
\rangle,\alpha\beta}\tilde{c}^{\dag}_{i,\alpha} {\mbox{\boldmath 
$\sigma$}}_{\bar{\beta}\bar{\alpha}} \tilde{c}_{j,\beta} {\bf S}_l ,
\end{eqnarray}
where $\langle ij \rangle_{2(3)}$ denotes the second (third)
next-nearest neighbor sites, $\langle ilj \rangle$ denotes the 
three nearest neighbor sites ($\langle il \rangle$, $\langle jl 
\rangle$), $\hat{N}_l$ is the number of fermions operator, 
${\bf S}_l=\frac{1}{2}c^{\dag}_{l,\alpha} {\mbox{\boldmath 
$\sigma$}}_{\alpha\beta}c_{l,\beta}$ is the local spin operator,
{\boldmath $\sigma$} is the Pauli matrix, $\bar{\alpha}=-\alpha$. 
The differences in the explicit form of the  
$\tilde{c}^{\dag}_{i,\alpha}(\tilde{c}_{i,\alpha})$ operators for the 
hole- and electron-doped systems as well as the signs of  
$t^{\prime}, t^{\prime\prime}, t^N, t^S$ will be discussed later. Now 
we will consider the origin of the $t^{\prime}, t^{\prime\prime}$ and 
$t^N, t^S$ terms.

Let us begin from the last ones. The second-order perturbation 
treatment of the usual Hubbard model near half filling provides the 
effective superexchange interaction (Eq. (\ref{1})) and the so called 
`three-site spin-dependent hopping' \cite{Psal1,Trug1,Eder1,Good2}. 
Their importance for both the spectrum of the charge excitation and 
the hole-hole interaction was noted \cite{Trug1}. These terms in 
Eq. (\ref{2}), manifestly, have the 
rotationaly invariant form. In the case of the usual Hubbard model
mapping to the $t$-$t^{\prime}$-$J$ model, the  expressions for the
$t^N, t^S$ terms have the simplest form:
\begin{eqnarray}
\label{3a}
t^{N}=-\frac{1}{2}t^{S}=\frac{t^2}{2U}=\frac{J}{8}. 
\end{eqnarray}
It should be noted that this expression (\ref{3a}) is valid for both 
types of doping, and that the signs of $t^N$ and $t^S$ 
are the same for the electron and hole due to 
the second-order nature of these terms. What would one expect from 
the three- or more-band Hubbard model mapping?

Much more high-energy channels (triplet, etc.) are opened for the  
virtual second-order transitions in comparison with the usual
Hubbard singlet only case \cite{Bel3}. This leads to two 
effects:  (i) $t^N\neq -\frac{1}{2}t^S$, and (ii) $t^{N, S}_h\neq 
t^{N, S}_e$, where $e$ and $h$ denote the electron and hole hopping 
integrals, respectively. The exact expressions for $(t^N)_{e, h}$ 
and $(t^S)_{e, h}$ were derived in Refs. \cite{Bel2,Bel3}. In spite 
of the contribution of the highest states, the lowest singlet is the 
 most important. The main features of these terms in the effective 
Hamiltonian (\ref{2}) remain unchanged, i.e.:  (i) 
$sign(t^N)=-sign(t^S)=+1$ both for the electron and hole, and (ii) 
$|t^N|\sim|t^S|\simeq J/4$, so that they are not negligible.

 In addition, one can point out that at the mean-field level, 
$\hat{N}$ 
 and $\bf{S}$ operators should be replaced with their averages 
$\langle 
\hat{N} \rangle$ and $\langle \bf{S} \rangle$, which effectively 
leads to:
\begin{eqnarray}
\label{4}
H_{MF}^{N,S}= \tilde{t}^{\prime} \sum_{\langle ij\rangle_2,\alpha}
\tilde{c}^{\dag}_{i,\alpha}\tilde{c}_{j,\alpha} + \tilde{t}^{\prime\prime} 
\sum_{\langle ij\rangle_3,\alpha} \tilde{c}^{\dag}_{i,\alpha}
\tilde{c}_{j,\alpha}  ,
\end{eqnarray}
with $\tilde{t}^{\prime}\equiv2\tilde{t}^{\prime\prime}=2 (\langle 
\hat{N} \rangle t^N + \langle {\bf S} \rangle t^S$), where the
additional factor 2 for the diagonal hopping integral 
$\tilde{t}^{\prime}$ as compared to the `oversite'   
$\tilde{t}^{\prime \prime}$ arises from the two possible ways for 
the virtual processes on the square lattice. Thus, the 
second-order terms in the Hamiltonian (\ref{2}) can be approximately 
considered as the renormalizations of the `bare' $t^{\prime}, 
t^{\prime \prime}$ amplitudes (\ref{4}).  As was noted, these 
renormalizations do not change the signs under changing of the doping 
type.

Now, return to the first two terms of the Hamiltonian (\ref{2}). They 
are the first-order terms arising in the first-principle 
models for the CuO$_2$ plane from the nonzero O-O hopping. While 
including of the $t^{\prime}$-term is evident 
\cite{Chub1,Dagrev,Lee1}, including of the $t^{\prime \prime}$-term 
requires an additional explanation. Physically, only the
$t^{\prime}$-term can arise for the next-nearest neighbor 
CuO$_4$-CuO$_4$ local states due to the $t_{pp}$ matrix element 
\cite{Esk1,Esk2}. However, as was shown 
\cite{Zh1,Jeff1,Bel3,Shas1}  the correct state of the oxygen 
low-energy degrees of freedom are the ortogonalized
Wannier states. Consequently, some 
`unphysical' transition amplitudes to the more distant neighbors 
arise.  These amplitudes fall rapidly with distance, and only the 
$t^{\prime}, t^{\prime \prime}$-terms should be kept in the 
low-energy model \cite{Bel3}. 

As was recently proposed in many works, the absolute sign of 
the $t^{\prime}$-terms for the electron- and hole-doped CuO$_2$ 
systems could be the source of the strong electron-hole asymmetry in 
the magnetic phase diagram \cite{Good1,Chub2}. 

The electron-hole asymmetry arises naturally in the three-band model 
\cite{Esk1,Bel2,Bel4,Bel3}. Not only the 
signs of the hopping integrals, but also the orders of the elementary 
processes, that lead to the effective hopping, differ for the 
electron and hole.  Consideration of a simple limiting case as well 
as discussion of the electron-hole asymmetry and  the dependence of
 the effective hopping integrals on the initial 
parameters are 
presented in Appendix A. Numerical results for the ranges of the 
effective parameters, based on the exact formulae of  
Ref.\cite{Bel3} and the approach developed in Ref.\cite{Bel4},  are 
shown in Table I.

Further, we will consider these (Table I) $t^{\prime}, 
t^{\prime\prime}, t^N, t^S$  values and $t_h=2.5J$, $t_e=-3.0J$ 
\cite{Bel2,Bel4} as the {\it realistic } ones. 

\section {Single-carrier energy spectrum}


Before doing the energy spectrum calculation let us discuss 
the sense of the constrained Fermi operators 
$\tilde{c^{\dag}}(\tilde{c})$ introduced in Eq. (\ref{1}). Both 
the hole and electron single-band low-energy Hamiltonians derived 
from the three-band Hubbard model are naturally expressed in terms of 
the Hubbard operators at the site $i$ 
\begin{eqnarray} 
\label{8a} 
X^{a b}_{i} \equiv |a\ i\rangle\langle b \ i|  , 
\end{eqnarray} 
here 
the states $a, b$ are either spin or singlet (vacancy) local states.  
Since the undoped state of the CuO$_2$ plane is set up by the 
localized holes in the Cu $d^{10}$ and O $p^6$ orbitals, the 
additional low-energy hole forms a singlet with the local one, 
whereas the electron in this background is a vacancy. Therefore,  
it is convenient to rewrite
 $t, t^{\prime}$-parts of Eqs.  (\ref{1}), (\ref{2}) as:  
\begin{eqnarray} 
\label{9} 
&&H_{hole}=\sum_{\langle ij 
\rangle_n, \alpha}t^h_{ij}X^{s\alpha}_i X^{\alpha 
s}_{j}+\sum_{\langle ilj\rangle,\alpha 
\beta}t_{h}^N X^{s\alpha}_{i}X^{\alpha s}_{j}\hat{N}_{l}  
\nonumber\\ 
&&\phantom{H_{hole}=} \mbox{}+\sum_{\langle 
ilj \rangle,\alpha \beta}t_{h}^S X^{s\alpha}_{i}X^{\beta 
s}_{j} ({\mbox{\boldmath $\sigma$}}_{\beta\alpha}{\bf 
S}_{l}) , \\ 
&&H_{el}=\sum_{\langle ij \rangle_n, 
\alpha}t^e_{ij}X^{0\alpha}_i X^{\alpha 
0}_j+\sum_{\langle ilj\rangle,\alpha 
\beta}t_{e}^N X^{0\alpha}_i X^{\alpha 0}_j\hat{N}_{l}  
\nonumber\\ 
&&\phantom{H_{hole}=} \mbox{}+\sum_{\langle ilj\rangle,\alpha 
\beta}t_{e}^S X^{0\alpha}_i X^{\beta 0}_j
({\mbox{\boldmath $\sigma$}}_{\beta\alpha}{\bf S}_l ) ,
\nonumber
\end{eqnarray}
where $|\alpha \rangle=|\uparrow \rangle, |\downarrow \rangle$ is a 
local spin state, $|s \rangle$ and $|0 \rangle$ are singlet and 
vacancy states, respectively, $\langle ij \rangle_n$ denotes 
all essential neighbor sites. Note, that the signs of the next-nearest 
hopping parameters shown in Table I are related to {\it these} 
particular (Eq. (\ref{9})) order of the Hubbard operators. The 
signs 
of the nearest neighbor hopping are:  $t^h_{\langle ij \rangle} > 0$, 
$t^e_{\langle ij \rangle} < 0$. The later is not essential, since for 
the antiferromagnetic background the shift of the quasimomentum space 
by the vector of the reciprocal lattice $Q=(\pi,\pi)$ changes the 
$sign(t_{\langle ij \rangle})$ but physically changes nothing 
\cite{Chub1}.

We wish to stress that when the kinetic-energy part is of the 
single-band type (as in Eq.(\ref{9})), mapping of the Hubbard 
operators onto the constrained Fermi operator basis is twofold.

If one prefers to retain the singlet ideology for the hole-doped 
system, the first term of the 
Hamiltonian (\ref{9}) can be rewritten identically as \cite{Sush1}:
\begin{eqnarray}
\label{10}
H_{hole}=\sum_{\langle ij \rangle_n, 
\alpha}t_{ij}^h 
n^h_{i,-\alpha}h_{i,\alpha}^{\dag}h_{j,\alpha}n^h_{j,-\alpha} , 
\end{eqnarray}
and the background is created by the $h_{i,\alpha}$-hole at every 
site.  $n^h_{i,-\alpha}=h^{\dag}_{i,-\alpha}h_{i,-\alpha}$ is the 
projection operator which project out the vacancy states (`double 
electron occupancy'). This representation for the electron over the 
hole background is:  
\begin{eqnarray} 
\label{11} 
H_{el}=-\sum_{\langle ij \rangle_n, 
\alpha}t_{ij}^e 
(1-n^h_{i,-\alpha})h_{i,\alpha}^{\dag}h_{j,\alpha}(1-n^h_{j,-\alpha}) , 
\end{eqnarray}
where the operators $(1-n_{i,\alpha})$ project out the 
double hole occupancy. An alternative way is to consider the physical 
hole as a `hole' (vacancy) in the upper Hubbard band 
\cite{Kane1,Plak1}, and the physical electron as the `particle' 
(singlet) in the lower one. This freedom in choosing is not connected 
with the initial orbital structure of the CuO$_2$ plane, but follows 
from the algebra of the Hubbard operators in Eq.(\ref{9}).  Hence, 
Eq.(\ref{11}) for an extra electron on the electron background can be 
written as:  \begin{eqnarray} \label{12} H_{el}=\sum_{\langle ij 
\rangle_n, \alpha}t_{ij}^e 
n^e_{i,-\alpha}e_{i,\alpha}^{\dag}e_{j,\alpha}n^e_{j,-\alpha} .
\end{eqnarray}
At the first glance, it would seem that the choice of the 
`particle' (Eq.(\ref{12})) or `vacancy' (Eq.(\ref{11})) language 
leads to the change of the sign of the quasiparticle energy. This 
is not true, since the `vacancy' energy has the sign of $t_{ij}$ 
reversed compared to the `particle' energy \cite{Chub1,Good2}.  
Therefore, Eqs. (\ref{11}) and (\ref{12}) lead to the {\it same} 
energy.

Thus, the explicit form of the constrained operators in Eqs. 
(\ref{1}), (\ref{2}) is: 
$\tilde{c}_{i,\alpha}=h_{i,\alpha}n^h_{i,\alpha} 
(e_{i,\alpha}n^e_{i,\alpha})$ for the physical hole (electron) 
system.  Hereafter, we will work with the Hamiltonian Eqs. (\ref{1}), 
(\ref{2}), constraint from Eqs. (\ref{10}), (\ref{12}), and 
parameters from Table I. 

Properties of the single-particle in an antiferromagnetic background 
were studied in detail by many authors 
\cite{Ba1,Ba2,Sush1,Auer1,Dagrev,Hors1,Ohta1,Manos1} using 
different approaches. Their results coincide at the point that the 
carriers are strongly dressed by the spin waves, i.e. quasiparticles 
are the magnetic polarons of a small radius with a strongly 
anisotropic dispersion law and small enough quasiparticle residue. We 
base our calculations on the results of Ref. \cite{Sush1}. 
The suggested 
trial function of an extra particle has the simple form \cite{Sush1}:  
\begin{eqnarray}
\label{13}
&&\psi^{\dag}_{{\bf k},\uparrow}=\frac{1}{\sqrt{N/2}}
 \sum_{n\in \downarrow} (\nu_{\bf k} c^{\dag}_{n, \uparrow}
\\
&&\phantom{\psi^{\dag}_{{\bf k},\uparrow}=} \mbox{}
+S^{\dag}_{n} \sum_{\delta}\mu_{{\bf k},\delta}  
c^{\dag}_{n+\delta, \downarrow}) \exp (i{\bf k}{\bf r}_{n}), 
\nonumber
\end{eqnarray}
valid also for $\psi^{\dag}_{{\bf k}, \downarrow}$ after 
changing $\uparrow \Leftrightarrow \downarrow$ and 
$S^+\Leftrightarrow S^-$.  $n\in \uparrow (\downarrow)$ means the 
site $n$ of the spin-up (-down) sublattice, $\delta$ is the unit 
vector to the nearest neighbor site. The explicit expressions for  
$\nu_{\bf k}$ and $\mu_{{\bf k}, \delta}$ are given in Appendix B. 
From the string picture point of view this simplest ansatz (\ref{13}) 
consists of the `bare' particle and four shortest `strings' of the 
lattice constant length. It was shown   
\cite{Sush1,Chern1,Sush2,Sush3} that the one-hole dispersion as well
as two-hole contact and long-range interactions, and even many-hole 
properties of the $t$-$J$ model are quantitatively well described 
using the above ansatz. 

Consideration of the  generalized $t^{\prime}$-terms (Eq.(\ref{2})) 
does not require any changes in the trial function (\ref{13}). 
Roughly, this is due to the following reasons: (i) 
the $t^{\prime}$-terms lead to the motion of the particle over  
{\it one} sublattice without distortion of the spin background, that 
enables `bare' particle  
to propagate freely; (ii) the motion of the `dressed' particle (with 
a string) mainly leads to longer strings, whose contribution 
to the energy is of the order of $\frac{t^{\prime}_{eff}}{t}$ and 
parametrically small at $t\simeq 3J$.  Hence, the expansion of 
the ansatz makes a small decrease in the energy, since the part of 
the ansatz (\ref{13}), which would be mainly affected by the 
$t^{\prime}$-Hamiltonian (\ref{2}), is the `bare' one.  The last
statement will be demonstrated below.

As was noted, the $t^{\prime}$-induced transitions occur in one 
sublattice, therefore a good preliminary consideration of the
role of each term in Hamiltonian (\ref{2}) can be done for the 
Ising background. It allows us to find easily the leading 
contribution to the band minima shifts, bandwidths, and 
effective masses. The $\hat{N}$ and ${\bf S}$ operators in 
(\ref{2}) should be replaced with $\langle \hat{N} \rangle =1$, 
$\langle 
{\bf S} \rangle=\langle S^z \rangle=\pm 1/2$. The eigenenergy of 
the magnetic polaron (\ref{13}) is 
\begin{eqnarray} 
\label{14} 
&&\epsilon_{\bf k}=\frac{\langle \psi_{\bf 
k}|(H_{t-J}+H_{t^{\prime}})| \psi_{\bf k} \rangle}{\langle \psi_{\bf 
k}| \psi_{\bf k} \rangle}\simeq E_0+\beta_1\gamma_{\bf 
k}^2+\beta_2(\gamma_{\bf k}^-)^2  ,
\nonumber\\ 
&&\beta_1=4\nu^2(t^{\prime}+2t^{\prime \prime}+4t^{N}-2t^{S}), 
\nonumber\\
&&\beta_2=4\nu^2(2t^{\prime \prime}-t^{\prime}), 
\\
&&\gamma_{\bf k}=1/2(\cos(k_x)+\cos(k_y)), 
\nonumber\\
&& \gamma^-_{\bf k}=1/2(\cos(k_x)-\cos(k_y)), 
\nonumber 
\end{eqnarray}
where we omitted all high-order ($t^{\prime}_{eff}/t$) terms.
E$_0$ is the depth of the band, $\beta_1, \beta_2$ are the inverse 
masses in the directions ortogonal and parallel to the magnetic 
Brillouin zone boundary, respectively. Since for the Ising case 
a particle of the pure $t$-$J$ model is dispersionless \cite{Bul1}, 
$\nu$ and $\mu$ are constants and at the realistic $t\simeq (2-3)J$ \ 
$|\nu|^2\simeq\frac{1}{2},\ |\mu|^2\simeq\frac{1}{8}$. 
Expression (\ref{14}) for the energy coincides almost exactly 
with  that obtained for the free 
spinless fermions \cite{Onuf1}. The main difference is the 
presence of the 
weight of the `bare' particle $|\nu|^2$ in Eq. (\ref{14}).
 In the realistic region of $t, t^{\prime}, t^{\prime \prime}$ 
etc $E_0$ is very close to its $t$-$J_z$ model value $E_0\simeq 
-2t$.

Two notes should be done. Firstly, the $\beta_1$-term 
(\ref{14}) represents the dispersion which is degenerate along the
$(\pi,0)-(0,\pi)$ line, whereas the $\beta_2$-term lifts out this 
degeneracy, placing the minima at the $(\pm \pi/2,\pm\pi/2)$ (if 
$\beta_2 > 0$) or $(\pm\pi,0),(0,\pm\pi)$ (if $\beta_2 < 0$) points.  
The difference $\Delta_{(0,\pi)-(\pi/2,\pi/2)}\equiv \beta_2$  is 
proportional {\it only} to the `first-order' terms $t^{\prime}, 
t^{\prime \prime}$ (\ref{14}) and does not depend on the second-order 
$t^N$, $t^S$ terms.  Secondly, one can see from Eq. (\ref{14}) that 
in the case $t^N > 0$, $t^S < 0$ and 
$sign(t^{\prime})=-sign(t^{\prime\prime})$  $\beta_1$ and bandwidth 
$W (=\beta_1$ or $\beta_1-\beta_2)$ are mostly determined by a few
$(2t^N-t^S)$.

The above qualitative calculation for the simple Ising background 
results in conclusions, which remain valid for the N\'{e}el case. The 
energy difference between the $(\pi/2,\pi/2)$ and $(\pi,0)$ points, 
which is the crucial value for the various calculations, has 
different signs for the hole- and electron-doped systems and is 
{\it not small } for both cases. It means that if the 
single-particle picture is valid for the finite doping regime, the 
quasiparticle Fermi-surface will be located (up to the high enough 
doping level) near the points $(\pm\pi/2,\pm\pi/2)$ and $(\pm\pi,0), 
(0,\pm\pi)$ for the hole- and electron-doped systems, respectively.

Also, it can be shown from Eq. (\ref{14}) that, at 
least for the upper limit of the $t^{\prime}, t^{\prime \prime}, t^S, 
t^N$ hopping parameters (Table I), the bandwidths are large enough. 
Thus, one would expect decrease of the DOS
compared to the pure $t$-$J$ model one. 

Finally, the role of the `bare' particle for the considered simple 
case is crucial, since the `dressed' part cannot  propagate freely
in the absence of the spin fluctuations. Therefore, the weight of the 
bare part in the trial wave function is the subject of the prime 
interest. As was noted $n^{bare}=\nu^2\simeq 0.5$, that is larger 
than in other works \cite{Hors1,Ohta1,Manos1}  where 
$\tilde{n}^{bare}=\tilde{\nu}^2\simeq 0.35-0.41$. For  $n^{1 
string}=4|\mu|^2\simeq 0.42$ the agreement is better: 
$\tilde{n}^{1 string}\simeq 0.44$. This discrepancy is due to the
transfer of the weight from the rest of the {\it exact} particle wave 
function (with the infinite number of strings) to the weight of 
the `bare' particle in the approximate ansatz (\ref{13}). Thus, our 
main approximation lies not in the `shortness' of the ansatz, but in 
the overestimation of the bare particle weight. This problem can be 
overcome by the simple renormalization of $\nu^2$ to $\tilde{\nu}^2$.

Let us discuss now what changes of the above results can be 
expected for the N\'{e}el background. The main changes arise from 
the fact that due to the spin fluctuations the pure $t$-$J$ model 
particle is given the possibility to propagate. Evidently, this will 
provide additions to $\beta_1, \beta_2$ and the bandwidth $W$. Also, 
the `dressed' part will lead to some coherent transitions. Due to the 
more complex structure of the matrix elements of the Hamiltonian 
(\ref{1}), (\ref{2}) for the N\'{e}el background,
the simple tight-binding relations between $\beta_1, 
\beta_2$ and $W$ no longer hold.

The pure $t$-$J$ dispersion law \cite{Sush1,Sush2} is:
\begin{eqnarray}
\label{17}
\epsilon^{t-J} ({\bf k})&=&2J
                  -\sqrt{(0.66J)^{2}+
4.56t^{2}-2.8 t^{2}\gamma^2_{\bf k}} 
\nonumber \\ 
& &+0.01 |t| (\gamma^{-}_{\bf k})^2 
 \\ 
& &  \simeq 
E^{t-J}_0+\beta^{t-J}_1\gamma_{\bf k}^2+
\beta^{t-J}_2(\gamma_{\bf k}^-)^2 , 
\nonumber    
\end{eqnarray}
where $\beta_1^{t-J}\simeq +0.65|t|$ and 
$\beta_2^{t-J}\simeq +0.01|t|$. Due to the ${\bf k}$-dependence of 
the trial wave function components $\nu_{\bf k},  \mu_{\bf k}$ and 
the nonzero matrix elements for the string components, the first 
order contribution to the energy from the $t^{\prime}$-Hamiltonian 
(\ref{2}) contains extra terms with the highest powers of 
$\gamma_{\bf k}^2$:  
\begin{eqnarray} 
\label{18} 
&&\epsilon ({\bf k})=\epsilon^{t-J} ({\bf k})+ 
\delta\epsilon^{t^{\prime}}({\bf k}) \\ 
&&\delta\epsilon^{t^{\prime}}({\bf k})=\delta 
E_0+\delta\beta_{1,{\bf k}}\gamma^2_{\bf k}
+\delta\beta_{2,{\bf k}}(\gamma^-_{\bf k})^2
\nonumber \\
&&\phantom{\delta\epsilon^{t^{\prime}}({\bf k})=} \mbox{}
+A_{\bf k}\gamma^4_{\bf k}+
B_{\bf k}\gamma^2_{\bf k}(\gamma^-_{\bf k})^2 ,
\nonumber
\end{eqnarray}
where the $\delta\beta_{1,{\bf k}}, \delta\beta_{2,{\bf k}}, A_{\bf 
k}$ and $B_{\bf k}$ coefficients weakly depend on ${\bf k}$ 
through the $|\nu_{\bf k}|^2$ and $|\mu_{\bf k}|^2$ quantities (see 
Appendix B).  As was discussed earlier, the highest-order 
corrections to the energy from Eq. (\ref{2}) are of the order of 
$\frac{t^{\prime}_{eff}}{t}$, and for the realistic 
$t/J\simeq(2-3)$ and $t^{\prime}$ from Table I can be neglected 
($\sim \frac{1}{10}$).  Since $|\nu_{\bf k}|^2$ is almost  
constant, the renormalization to the `normal' $|\tilde{\nu}|^2$ can 
be accomplished by the simple replacing $|\nu_{\pi/2, 
\pi/2}|^2\simeq|\nu_{\pi,0}|^2\simeq|\nu_{0,0}|^2 
\Rightarrow|\tilde{\nu}|^2\simeq0.35$.

Figures \ref{fig1}-\ref{fig4}  present our results for the hole and 
electron dispersions. Contour plot in Fig.\ \ref{fig1}
 shows the characteristic 
feature of the hole dispersion in the whole Brillouin zone. The 
magnetic Brillouin zone boundary as well as the $\Gamma 
(0,0)\rightarrow M (\pi/2,\pi/2) \rightarrow Z (\pi, \pi) \rightarrow 
Y (0, \pi) \rightarrow \Gamma$ and $Y \rightarrow M \rightarrow X 
(\pi, 0)$ directions are shown.  Figs. \ \ref{fig2}-\ref{fig4} show 
the dispersions along the $\Gamma \rightarrow M \rightarrow Z 
\rightarrow Y  \rightarrow \Gamma$, and $Y \rightarrow M \rightarrow 
X $ lines.  Fig.\ \ref{fig2} demonstrates the characteristic 
$t$-$t^{\prime}$-$J$ hole dispersion (solid line) for the average 
parameters from Table I, pure $t$-$J$ model hole dispersion 
(dashed line), and pure $t^{\prime}$-$J$-dispersion $t\equiv 0$ 
(dotted line).  This qualitative picture clearly demonstrates that 
due to the $t^{\prime}$-terms the system becomes `less strongly 
correlated' because of the `liberating' of the bare carrier. Figs.\ 
\ref{fig3} and \ref{fig4} show dispersions along the same lines.  
Fig.\ \ref{fig3} describes the hole-doped system, and Fig.\ 
\ref{fig4} the electron-doped one. Solid curves are related to the 
upper and lower limit for the $t^{\prime}, t^{\prime \prime}, t^N, 
t^S$ parameters from Table I, the dotted satellites are related to 
the same parameters for the renormalized bare carrier weight (up to 
0.35). The dashed lines demonstrate the pure $t$-$J$ model 
dispersions.

Our calculations of the $t^{\prime}$-part (\ref{18}) in the 
dispersion demonstrate that the `bare' particle contribution for the 
N\'{e}el case is very close to the results for the Ising background 
and that the `dressed' particle contribution plays a minor role 
($<0.2$) compared to the `bare' one.  Using the data from Table I, the 
variations of the inverse mass
 values and the bandwidths for the hole and 
electron are found as follows:
\begin{eqnarray}
\label{19}
&&\beta^h_1=(3.8 - 5.3)J, \ \ \ \ \beta_1^e=(3.5 - 5.0)J, 
\nonumber \\ 
&&\beta^h_2=(1.3 - 3.1)J, \ \  
\beta^e_2=-(0.7 - 1.6)J, \\ 
&&W^h=(5.1 - 7.1)J, \ \ \ 
W^e=(5.2 - 8.4)J, 
\nonumber 
\end{eqnarray}
In spite of some changes in the inverse mass values (\ref{19}), 
the features discussed earlier for the Ising case are still valid. 
Thus, the shifts of the band minima are rather large, the bandwidths 
are substantially wider as compared to those in the $t$-$J$ model, 
and the role of the bare particle weight remains the most 
important.

One of the unsolved problem of the $t$-$J$ type 
models is: whether the free-particle approximation work for the doped 
systems \cite{Good2,Ohta1,Hors2}.  Therefore, the extension of the 
single-particle calculation onto the case of finite doping 
indeed requires justifications.  Some of them can be
found in Refs.  \cite{Dagrev,Good2,Ohta1}. We simply accept the 
free particle picture and have calculated the chemical 
potential as the function of concentration for the dispersion law 
(\ref{17}), (\ref{18}). From Figs. \ \ref{fig1}-\ref{fig4} it is 
clear that the ground state is $(\pm\pi/2, \pm \pi/2)$ for the 
hole, and $(0,\pm\pi)$, $(\pm\pi,0)$  for the electron.  Both minima are 
not shallow:  $|\Delta_{(0,\pi)-(\pi/2,\pi/2)}| \sim J$.  At the same 
time, the bandwidths are larger than $t$-$J$ ones 
($W^{t-J}\sim 2J$), and the accumulation of the DOS at low 
energy is smaller than for the $t$-$J$ model case. This point may be 
of interest in view of intensive discussion of the possible 
Van Hove singularity at the optimal doping \cite{DagVH}.  The above 
mentioned two different tendencies to increase the minima shifts and 
to decrease the DOS require a quantitative consideration.  Figs. \ 
\ref{fig5} and \ref{fig6} show our results for the DOS versus 
energy $E/J$ and versus  concentration $\delta$, 
respectively. One can see that in spite of the DOS lowering it 
remains rather large at low energies (since $(3-8)J$ $\ll 8t$), 
and the peaks in the DOS shift to higher concentrations 
($\delta_{max} \sim 0.4-0.5$), compared to the $t$-$J$ case 
($\delta_{max} \sim 0.1-0.2$). 

\section {Conclusions}


To summarize, in this paper we have studied the microscopically 
derived extended type of the $t$-$t^{\prime}$-$J$ model.  Previous 
investigations of the three-band model allowed us to establish the 
necessary next-neighbor terms for the low-energy single-band model. 
In this work the efficient scheme, developed previously for the 
calculation of the $t$-$J$ model parameters of the real CuO$_2$ 
systems, has been applied to the effective 
$t^{\prime}$-terms calculation.  This has enabled us to determine 
numerical ranges for all $t^{\prime}$-terms both for the electron- 
and hole-doped systems.

We also have performed simple calculations of the single-particle 
dispersion using the variational approach to the spin-polaron 
problem.  Both types of doping have been considered within the above 
mentioned {\it realistic } values of the $t$, $t^{\prime}_{eff}$ 
parameters. The importance of {\it all }  $t^{\prime}_{eff}$-terms 
has been clearly demonstrated: the $t^N$, 
$t^S$ terms of the second-order origin are responsible for the 
widening of the bandwidth, and the $t^{\prime}$, $t^{\prime\prime}$-terms
arising from the O-O hopping bring about the shift of the 
groundstate minima.  It has been found that the minima shifts are 
large enough ($\sim J$) and have opposite signs for the hole and 
electron systems. Also, the bandwidths have been found to be 
$1.5-4.0$ times larger than those in the pure $t$-$J$ model.

An investigation of the finite doping regime for the simple 
free-particle approximation has been carried out. The Van Hove peak 
in the DOS has been found to decrease and move to a higher doping 
level from its $t$-$J$ model position.

\vskip 1.cm

{\Large \bf Acknowledgments} \\ \vskip 0.1cm
It is our pleasure to thank A. Barabanov, L. Maksimov and O. 
Sushkov for initiative discussions and M. Avignon for helpful 
conversation.
We are grateful to O. A. Starykh for instructive discussion and 
sending their results to us prior to publication.
This work 
was supported in part by the  Russian Foundation for 
Fundamental Researches, Grant No 94-02-03235; International 
Science Foundation established by G. Soros, Grant No J2M100; The 
Competition Center for Natural Sciences at St.-Peterburg State 
University, Grant No 94-5.1-1060; Council on Superconductivity of 
Russian Academy of Sciences, Grant No 93197;
by the Scientific-Technical Program "High - Temperature Superconductivity"
as part of the state program "Universities as Centers for Fundamental
Research".
 The work of one of us 
(A.L.C.) has been supported by a fellowship of INTAS Grant 93-2492 
and is carried out within the research program of International 
Center for Fundamental Physics in Moscow.


\appendix
\section{}

It will be useful for the future discussion to consider the simple 
limiting case of the three-band model, when the Cu-O hopping is much 
less than both the Cu-Cu Coulomb repulsion U$_d$ and Cu-O levels 
splitting $\Delta$ ($t_{pd}\ll$ U$_d$,$\Delta$). 

The additional hole moves over 
the spin background as a Zhang-Rice (ZR) singlet \cite{Zh1}, whereas 
the 
electron moves as a vacancy. Corresponding hopping integrals in the 
lowest order of $t_{pd}/\Delta$ are:
\begin{eqnarray}
\label{a1}
&&t_h=0.53 \frac{t_{pd}^2}{\tilde{\Delta}}+0.275 t_{pp},\ \
 t_e=-(1.06+2.03 \frac{t_{pp}}{\tilde{\Delta}})
\frac{t_{pd}^2}{\tilde{\Delta}}, 
\nonumber\\
&&t_h^{\prime}=0.09 \frac{t_{pd}^2}{\tilde{\Delta}}- 0.124 t_{pp}, \ 
\ t_e^{\prime}=-(0.18-0.92\frac{t_{pp}}{\tilde{\Delta}}) 
\frac{t_{pd}^2}{\tilde{\Delta}},
\nonumber\\
&&t_h^{\prime \prime}=0.053 
\frac{t_{pd}^2}{\tilde{\Delta}} + 0.062 t_{pp}, \ 
t_e^{\prime\prime}= -
(0.105 + 0.46 \frac{t_{pp}}{\tilde{\Delta}}) \frac{t_{pd}^2} 
{\tilde{\Delta}},
\nonumber\\
&&\tilde{\Delta}=\Delta-1.45 t_{pp} .
\end{eqnarray}
The first two expressions for the hole hopping integrals coincide 
with those in the work by Jefferson {\it et al} 
\cite{Jeff1}, except the signs. The signs are opposite for the 
following reasons.

In calculation of the transfer amplitudes of the $t$-$J$ model 
singlet or vacancy from the three-band model, it is convenient to 
construct the basis of the wave functions as the direct product of 
the states at the sites: $(\hat{A}_1^{\dag}|0 \rangle_1 \otimes 
\hat{A}_2^{\dag}|0 \rangle_2 \otimes \ldots )$. The $\hat{A}_i^{\dag}$
operators can be considered as the Hubbard 
operators at the site $i$, which create the singlet or spin state 
over the vacuum state $|0 \rangle_i $.  Namely, the two-site wave 
function of the form $|s \rangle_i \otimes |\uparrow \rangle_j $ has 
been used for the calculation of the singlet ($|s \rangle_i$) hopping 
integrals (see Refs.  \cite{Esk1,Jeff1,Bel3}). This convenience is 
due to the complex structure  of the three-band low-energy states. 
For example, ZR singlet consists of the linear combination of a  
copper singlet (double copper occupancy), an oxygen one (double 
oxygen occupancy), and a copper-oxygen one. This state is hardly 
expressed in the terms of the Fermi creation operator which acts on 
the spin background.  However, the use of the 'direct-product' basis 
is incorrect when more than one Hubbard $\hat{A}_i^{\dag}$ operators  
are of the Fermi type.  This is evident from their anticommutativity.  
In the other words, the wave function of the half-filled background 
(one fermion per site) cannot be uniquely determined in this basis.  
Formally, the two-site wave function $|s \rangle_i \otimes |\uparrow 
\rangle_j$ is determined correctly since the singlet $|s \rangle_i$ 
is the boson.  Nevertheless, for the matrix elements of the hopping 
 Hamiltonian, which consists of the creation and annihilation 
operators of an additional fermion, one gets:  \begin{eqnarray} 
\label{a2}
&& t_{ij} \langle 
\uparrow|_i\otimes \langle s|_j \biggl[ \bigl(|s \rangle \langle 
\uparrow|\bigr)_j\cdot \bigl(|\uparrow \rangle \langle s|
\bigr)_i \biggr] |s 
\rangle_i \otimes |\uparrow \rangle_j
\nonumber \\
 \equiv&& t_{ij} \langle 
\uparrow|_i\otimes \langle \uparrow|_j \cdot |\uparrow \rangle_i 
\otimes |\uparrow \rangle_j.
\end{eqnarray} 
Thus, the matrix element reduces to the projection of the one 
half-filled state to the other and hence has an uncertain sign.  
Therefore, to avoid this uncertainty one has to use a more 
conventional basis, namely:  $c_i^{\dag}| gs \rangle$, where $| gs 
\rangle$ means the groundstate (one fermion per site), and 
$c_i^{\dag}$ is the creation operator of the additional fermion.  
This careful approach leads to the signs of the hopping amplitudes as 
presented in Eq.(\ref{a1}) and in Table I.  Note that this 
difference of the bases is absent for the single fermion in the 
lattice or for particles obeying Bose statistics.
  
 To obtain Eq. (\ref{a1}) we have used the 
general expressions for $t_{ij}$ from our previous works
\cite{Bel2,Bel3}. 

Approximate expressions (\ref{a1}) for $t$, $t^{\prime}$ and 
$t^{\prime \prime}$ are good enough for large and even moderate 
$\Delta$.  Thus, our calculations of the $t$,  $t^{\prime}$, 
$t^{\prime \prime}$ realistic values  for the CuO$_2$ plane,
show that $|t^{\prime}| < |t^{\prime \prime}|$ at least 
for small values of $V_{pd}$ (Cu-O Coulomb repulsion). This is due to 
the partial compensation of the Cu-O and O-O contribution
to the $t^{\prime}$-term, that can be seen in 
Eq.(\ref{a1}).  It should be noted that the $t_{pp}$  contribution to 
the hopping of the vacancy arises only in the third order of 
$t_{pd}/\Delta$ ($t_{pp}/\Delta$). The above mentioned compensation 
of the O-O and Cu-O amplitudes for the electron $t_e^{\prime}$-term 
is even more pronounced than for the hole. The exact formulae 
\cite{Bel2} provide very small $t_e^{\prime}$ for $V_{pd}=0$.

 To be more specific, in our 
calculations we have followed the idea of the 
narrowing of the uncertainty region for the low-energy model 
parameters using some experimentally observed values. This approach 
was developed in Ref.\cite{Bel4} and enabled us to calculate 
$t_h$ and $t_e$ for the real CuO$_2$ planes. Our results for the
$t^{\prime}, t^{\prime \prime}, t^N, t^S$ hopping integrals obtained 
in the same way for the two types of doping are shown in Table I. 

We wish to stress the strong $V_{pd}$-dependence of $t^{\prime}$ 
both for the hole and electron carriers. This 
 $t^{\prime}(V_{pd})$ dependence can be easily 
understood.  The rising of $V_{pd}$ requires the lowering of the 
$\Delta$ for the fixed values of the observable quantities. Thus, the 
occupation number of oxygen sites also rises, which immediately leads 
to the growing of the $t_{pp}$ contribution.  Note, that due to the 
smaller role of the oxygen degrees of freedom for the electron, its 
hopping integrals are less varying than those for hole.  This is in 
agreement with the cluster calculations by Eskes and Sawatzky 
\cite{Esk1,Esk2}.  In our three-band model calculations we used the 
experimental values of $J=0.14$ and $J=0.17$ eV (lanthanum and 
yttrium systems).

\section{}

Explicit expressions for the components $\nu_{\bf k}$, $\mu_{\bf k}$ 
\cite{Sush1} are:
\begin{eqnarray}
\label{b1}
 \nu_{\bf k}&=& {1\over 2} \biggl[ {{\Delta_0+2S_{\bf 
k}}\over{XS_{\bf k}}} \biggr]^{1/2}, 
  \\ 
  \mu_{{\bf k},{\delta}} &=&  
  \mu^1_{\bf k} +  \mu^2_{\bf k} \gamma_{\bf k} e^{i{\bf k}{\delta}} 
\nonumber\\
  &=& {t\over {[ Y S_{\bf k} (\Delta_0+2S_{\bf k}) ]^{1/2} }}
   [(1+v) - (u+v)\gamma_{\bf k} e^{i{\bf k}{\delta}}]
\nonumber
\end{eqnarray}
where parameters $X$, $Y$, $u$, and $v$ are expressed in 
  terms of ground state correlators \cite{Sush1} and for the N\'{e}el 
state are: $\Delta_0=1.33J$, $X=0.8$, $Y=0.72$, $u =0.42$,  $v=0.12$.  
$S_{\bf k}\equiv \epsilon^{t-J} ({\bf k})$ is the pure $t$-$J$ model 
  dispersion (\ref{17}). Since $S_{\bf k}\sim |t|$ at $t>J$, 
  $\nu_{\bf k}$ and $\mu^1_{\bf k}$, $\mu^2_{\bf k}$ are  
  weakly varying functions of ${\bf k}$.
  
Explicit expressions for the $\delta\beta_{1,{\bf k}}$,  
  $\delta\beta_{2,{\bf k}}$, $A_{\bf k}$, $B_{\bf k}$ coefficients 
(Eq.(\ref{18})) are:
\begin{eqnarray}
\label{b2}
\delta\beta_{1,{\bf k}}&=& \nu_{\bf k}^2 C_1^1+ (\mu^1_{\bf k})^2 
C_1^2+ (\mu^2_{\bf k})^2 C_1^3 +\mu^1_{\bf k}\mu^2_{\bf k} 
C_1^4 ,
\nonumber \\ 
\delta\beta_{2,{\bf k}}&=& \nu_{\bf k}^2 C_2^1+ (\mu^1_{\bf k})^2 
C_2^2+ (\mu^2_{\bf k})^2 C_2^3 +\mu^1_{\bf k}\mu^2_{\bf k} 
C_2^4 ,
\\ 
A_{\bf k}&=& \nu_{\bf k}^2 C_3^1+ (\mu^1_{\bf k})^2 
C_3^2+ (\mu^2_{\bf k})^2 C_3^3 +\mu^1_{\bf k}\mu^2_{\bf k} 
C_3^4 ,
\nonumber \\ 
B_{\bf k}&=& \nu_{\bf k}^2 C_4^1+ (\mu^1_{\bf k})^2 
C_4^2+ (\mu^2_{\bf k})^2 C_4^3 +\mu^1_{\bf k}\mu^2_{\bf k} 
C_4^4 ,
\nonumber 
\end{eqnarray}
where 
\begin{eqnarray}
\label{b3}
&&C_1^1= 3t^{\prime}+6 t^{\prime\prime}+12t^N-8.4t^S , 
\nonumber \\ 
&&C_1^2= -5.8t^{\prime\prime}-5.8t^N-1.6t^S ,
\nonumber \\ 
&&C_1^3= -8.2t^{\prime\prime}-8.2t^N+2.9t^S , \ 
\nonumber \\ 
&&C_1^4= 9.8t^{\prime} -6.7t^{\prime\prime}+12.8t^N+3t^S ,
\nonumber \\ 
&&C_2^1= -3t^{\prime}+6 t^{\prime\prime} , \ 
\nonumber \\ 
&&C_2^2= -1.5t^S , \ C_2^3=C_2^4=C_3^1=C_4^1=0 ,
 \\ 
&&C_3^2= 5.8t^{\prime}+11.5 t^{\prime\prime}+23t^N-12.2t^S , \ 
\nonumber \\ 
&&C_3^3= 10.6t^{\prime}+16.4t^{\prime\prime}+37.7t^N-15.1t^S ,
\nonumber \\ 
&&C_3^4= 11.5t^{\prime}+23t^{\prime\prime}+23t^N-30.4t^S , 
\nonumber \\ 
&&C_4^2= -5.8t^{\prime}+11.5 t^{\prime\prime}+4.6t^S , \ 
\nonumber \\ 
&&C_4^3= -10.6t^{\prime}+16.4t^{\prime\prime}-4.9t^N+3.5t^S ,
\nonumber \\ 
&&C_4^4= -11.5t^{\prime}+23t^{\prime\prime} , 
\nonumber 
\end{eqnarray}
Numbers in Eq. (\ref{b3}) result from $\langle S^z_i 
S^z_j\rangle$, $\langle S^+_i S^-_j\rangle$ and other spin 
correlators for various neighbor sites.


\vskip .2cm
{\Large \bf Figure captions} \\ \vskip 0.1cm
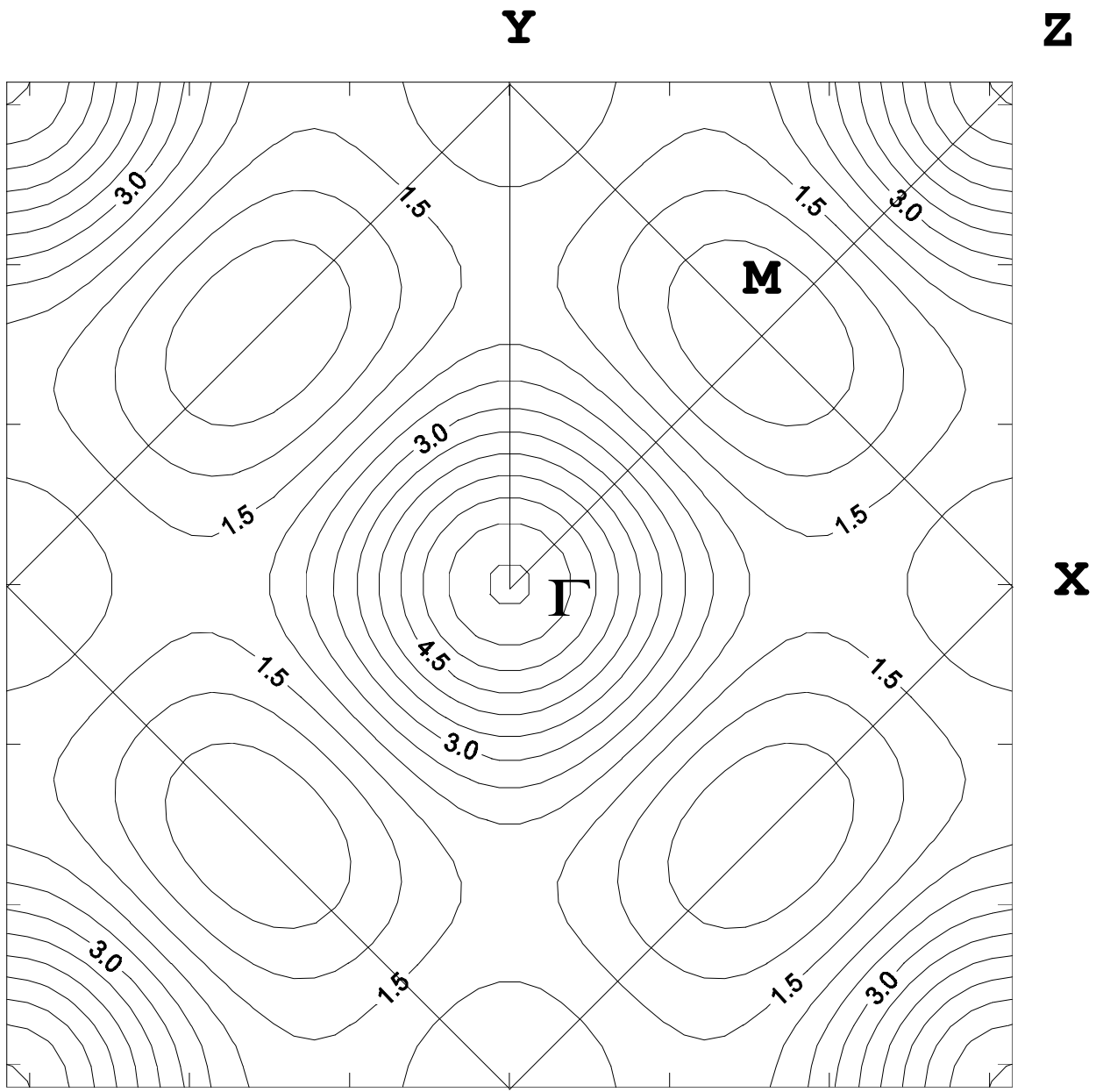
\begin{figure}
\caption{ Contour plot of  $\epsilon({\bf k})$ 
 for a hole. $t_h/J=2.5$, $t^{\prime}, 
t^{\prime\prime}, t^N, t^S $ are taken as the average values of the 
upper and lower limit from Table I. Inner square is the magnetic 
Brillouin zone boundary.  $\Gamma (0,0), M (\pi/2,\pi/2), Z (\pi,\pi),
Y (0,\pi)$, and $X (\pi,0)$ points are indicated. }
\label{fig1}
\end{figure}
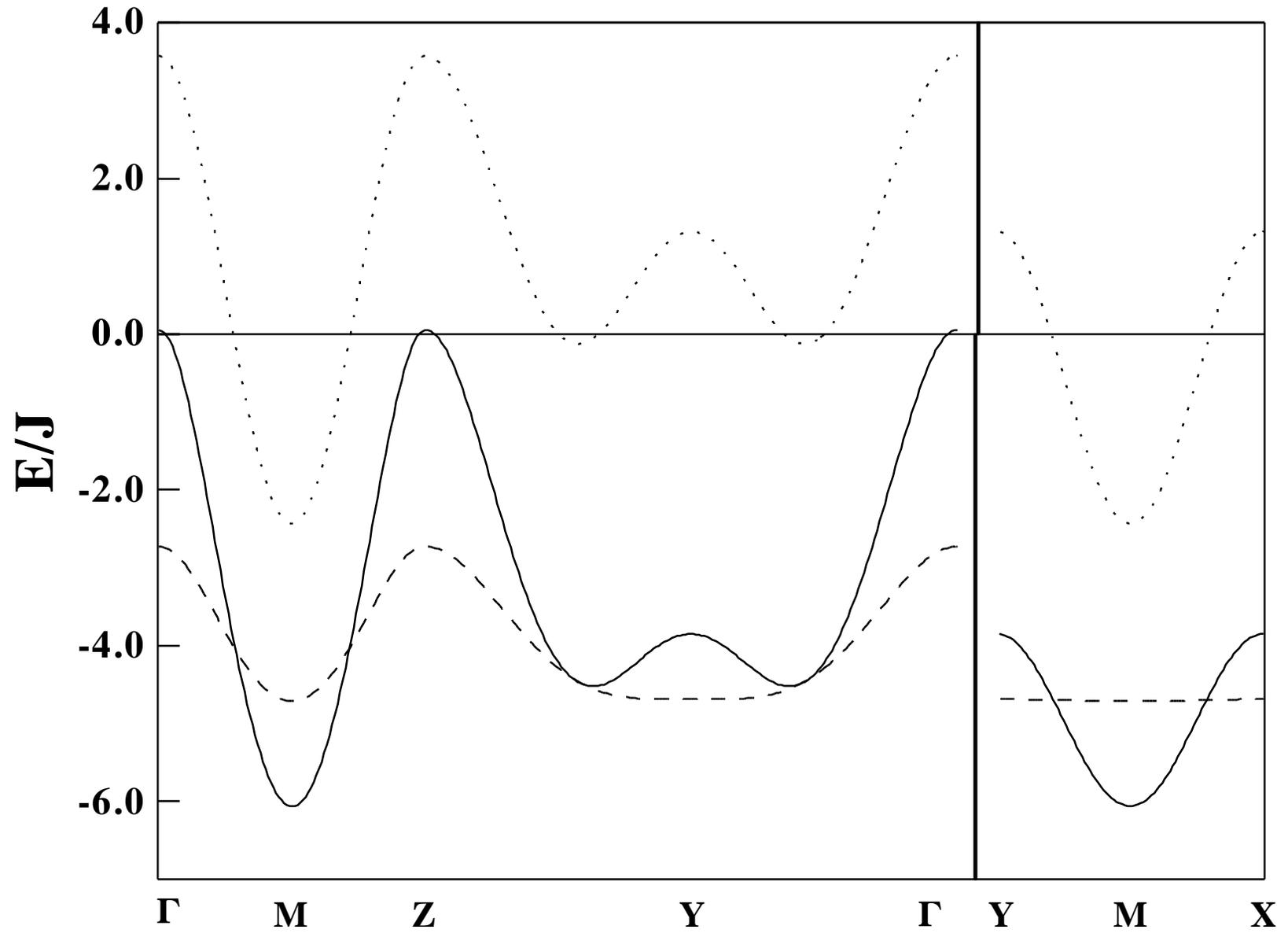
\begin{figure}
\caption{
Dispersion curves along the lines $\Gamma \rightarrow M 
\rightarrow Z \rightarrow Y  \rightarrow \Gamma$, and $Y \rightarrow 
M \rightarrow X$. Solid curve is the  $t$-$t^{\prime}$-$J$ hole 
dispersion for the same parameters as in Fig. 1,  
dotted curve is the pure $t^{\prime}$-$J$ dispersion ($t\equiv 0$), 
dashed curve is the pure $t$-$J$ model dispersion.}
\label{fig2}
\end{figure}
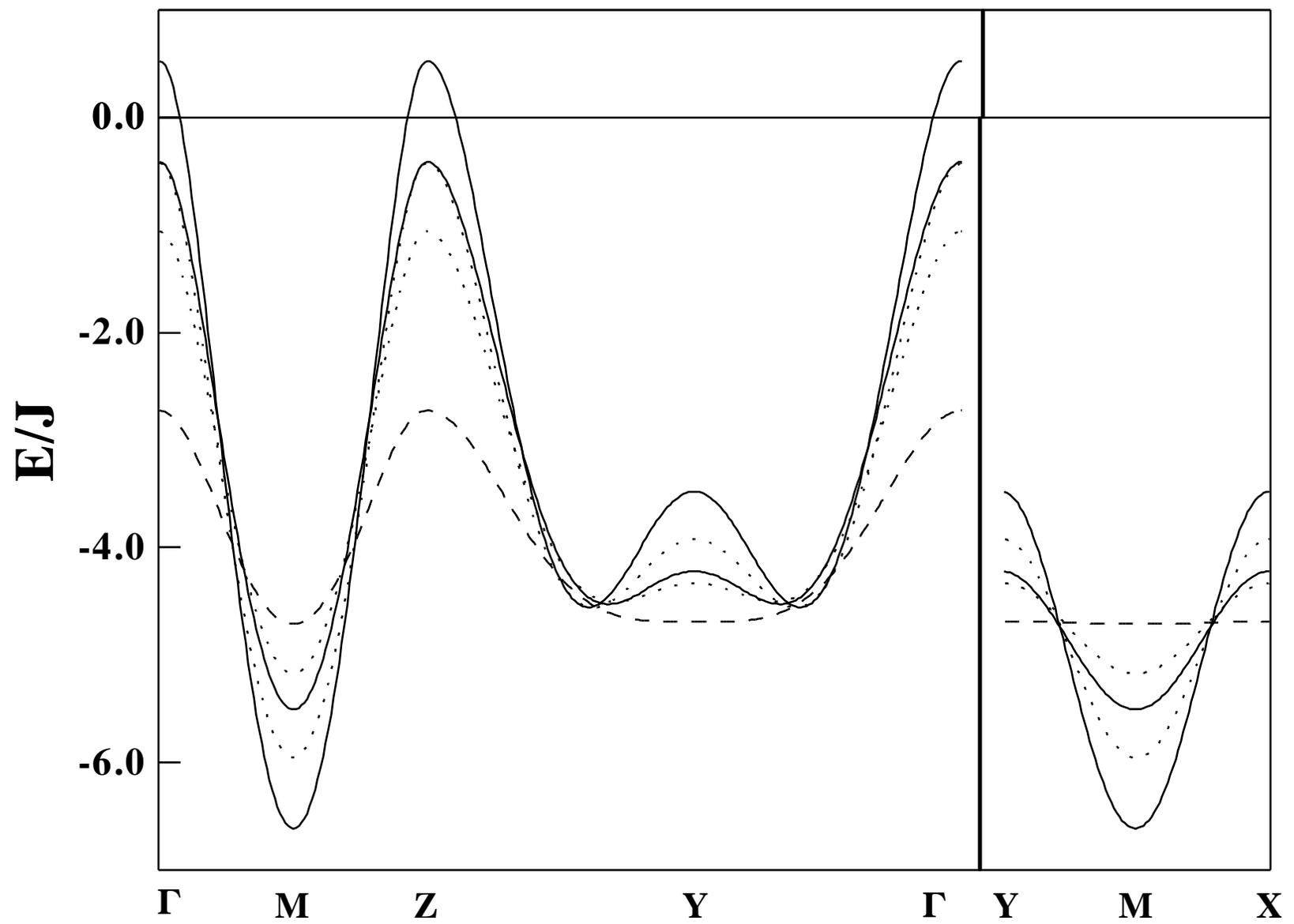
\begin{figure}
\caption{Dispersion curves along the lines as in Fig. 2 for a
hole.  Solid curves correspond 
to the upper and lower limits of the $t^{\prime}$-parameters (Table 
I).  Dotted curves correspond to the same with renormalized bare 
carrier weight. Dashed curve is the pure $t$-$J$ model dispersion.} 
\label{fig3}
\end{figure}
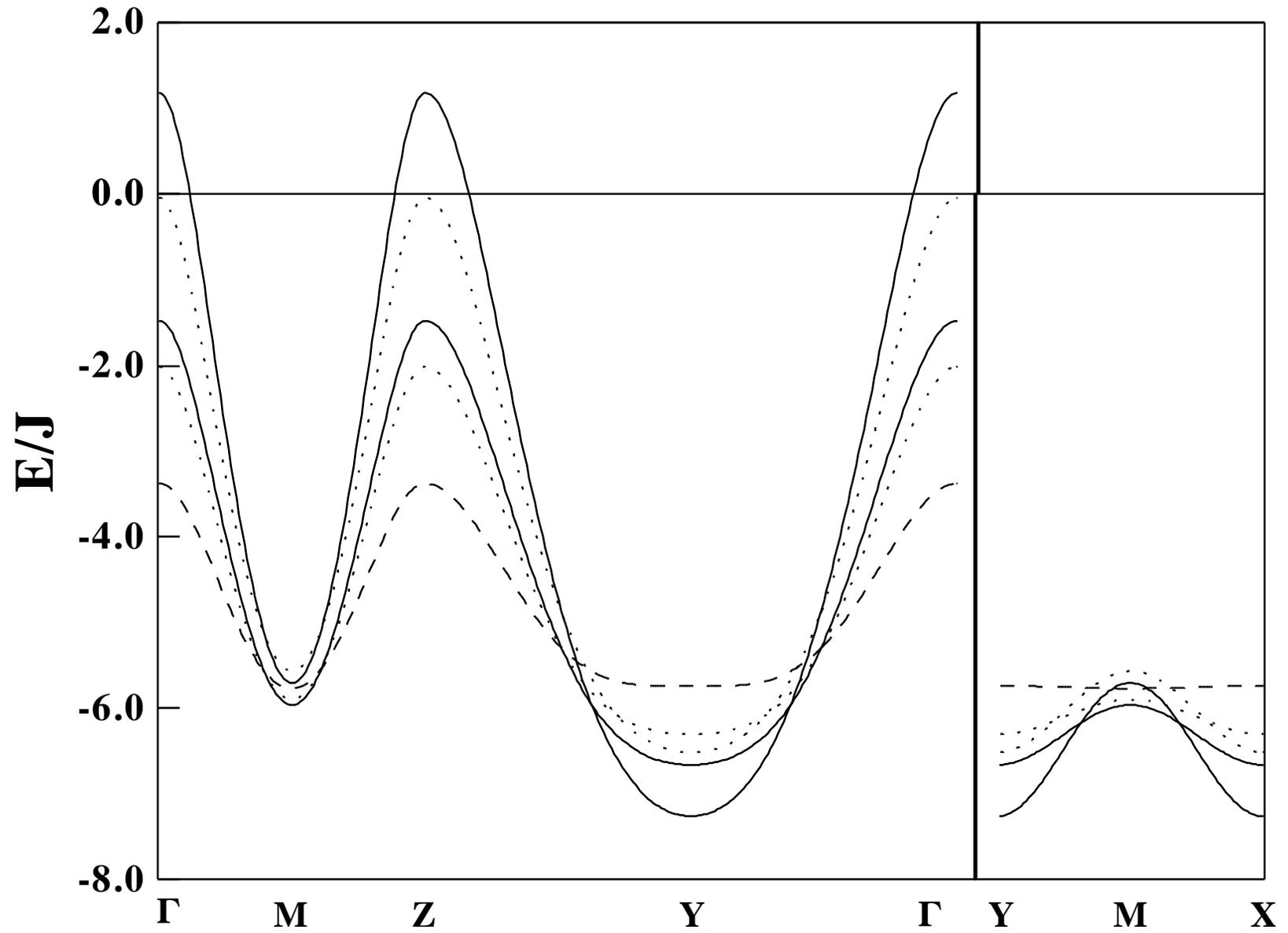
\begin{figure}
\caption{Dispersion curves along the same lines as in Fig. 2 for an
electron. All notations as in Fig.3.}
\label{fig4}
\end{figure}
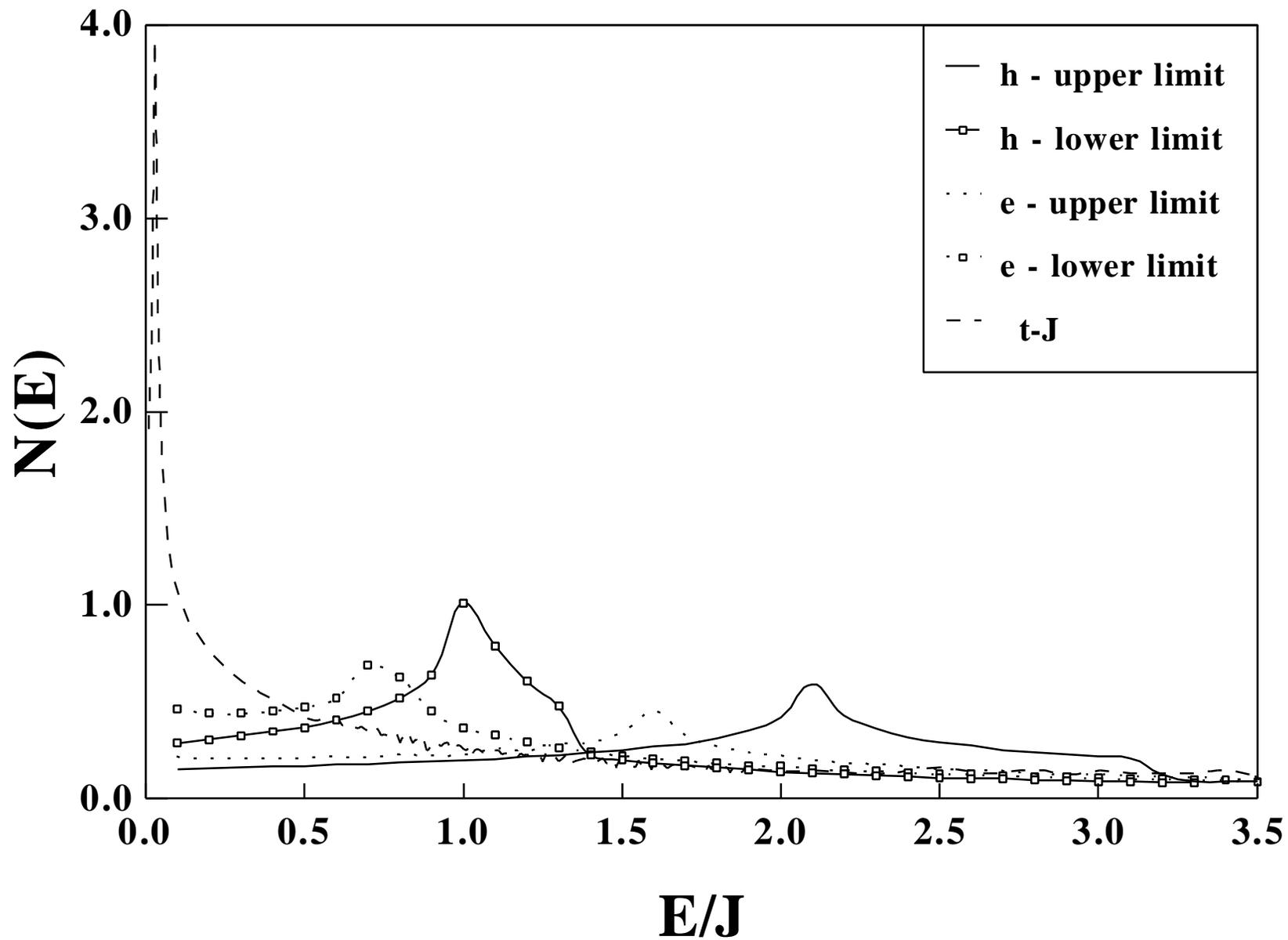
\begin{figure}
\caption{DOS {\it vs} energy $E/J$ for the pure 
$t$-$J$ model and electron and hole $t$-$t^{\prime}$-$J$ model (upper 
and lower limits in Table I). }
\label{fig5}
\end{figure}
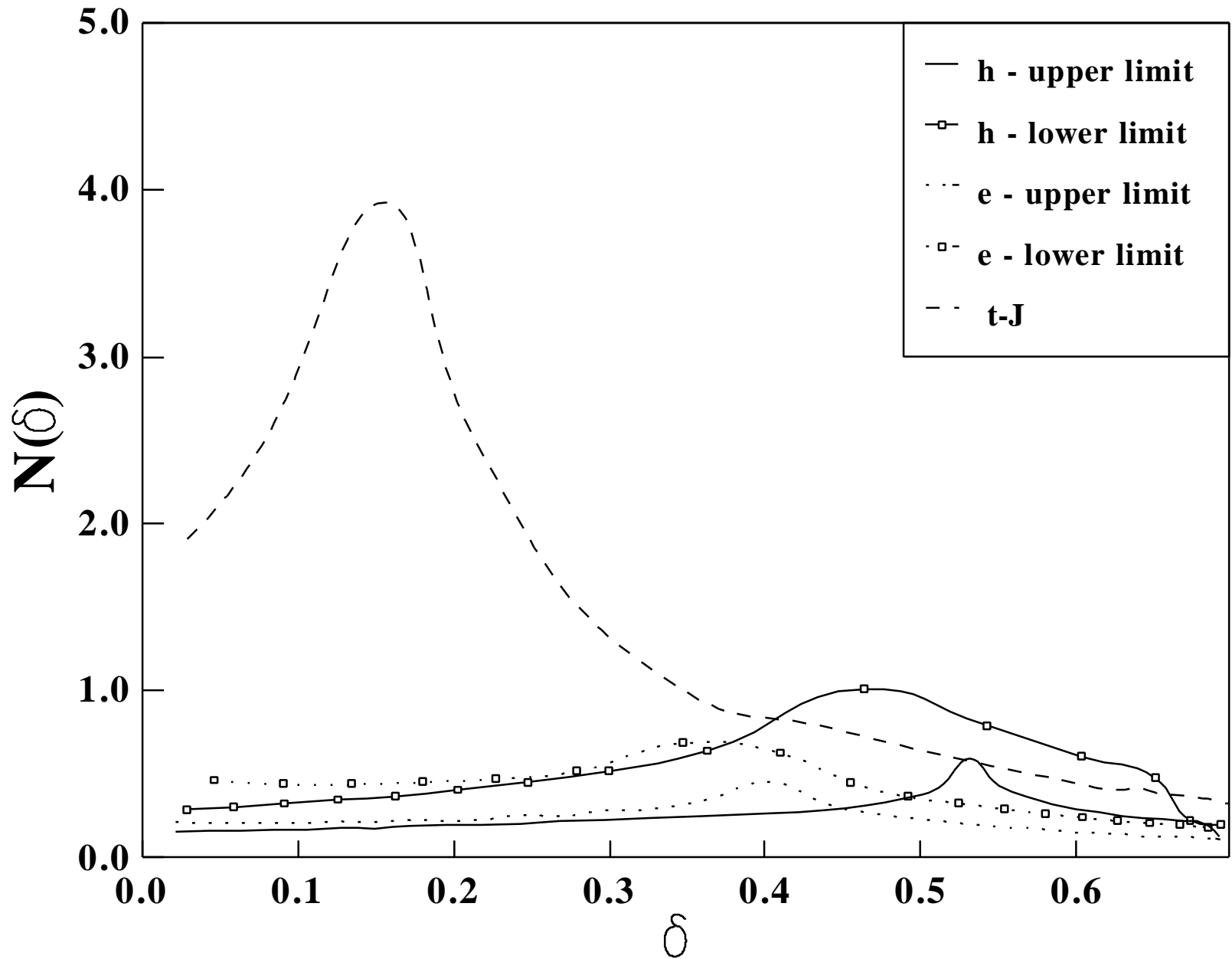
\begin{figure}
\caption{DOS {\it vs} concentration $\delta$.}
\label{fig6}
\end{figure}

\vskip .5cm
{\Large \bf Tables} \\ \vskip 0.1cm

Table I.  \ \ {\small $t^{\prime}, t^{\prime \prime}, t^S, 
t^N$ hopping parameters for the hole and electron. 
First and second rows show the lower and upper limits, respectively.} 
\vskip .5cm

\begin{tabular}{c|cccc}          
\hline   \\ 
& \ \ \ \ $t^{\prime}/|t|$\ \ \ & \ \ \ $t^{\prime 
\prime}/|t|$\ \  \ & \ \ \ $t^{N}/|t|$ \ \ \ &\ \ \ $t^{S}/|t|$\ \ \ 
 \ \\ \\ 
\hline  \\ 
hole & 0.01 & 0.12 & 0.01 & -0.07\\ 
$t_h/J=2.5$\ \ \ & -0.25 & 0.16 & 0.07 & -0.16 \\ \\ 
\hline \\  
electron\ & -0.09 & -0.11 & 0.07 & -0.10\\ 
$t_e/J=-3.0$\ & 0.03 & -0.12 & 0.11 & -0.12 \\  \\ 
\hline 
\end{tabular} 

\end{document}